\documentclass{aa}     
\usepackage{graphicx}
 
\def\cc{\,{\rm cm^{-3}}}
\def\cm2{\,{\rm cm^{-2}}}
\def\kms{\,{\rm {km\,s^{-1}}}}
\def\kkms{\,{\rm {K\,km\,s^{-1}}}}

\def\cmk{\,{\textstyle{\cm2 \over \kms}}}
\def\h2{\,{\rm H_{2}}}
\def\13co{\,{\rm ^{13}CO}}
\def\co{\,{\rm ^{12}CO}}

\def\half{\textstyle{1\over2}}
\def\threehalf{\textstyle{3\over2}}
\def\nco{\textstyle{N(CO)\over dV}}
\def\aua{{\rm A\&A} }

\def\apj{{\rm ApJ} }
\def\aj{{\rm AJ} }
\def\apjs{{\rm ApJS} }
\def\apjl{{\rm ApJL} }

\def\mnras{{\rm MNRAS} }

\def\pasp{{\rm PASP} }

\begin{document} 
 
\title{Dust and molecules in the Local Group galaxy NGC 6822} 
 
\subtitle{III. The first-ranked HII region complex Hubble~V} 
 
\author{F.P. Israel,
	\inst{1},
        F. Baas 
        \inst{1,2}\,\dag,
        R.J. Rudy
	\inst{3},
        E.D. Skillman
	\inst{4}
        and C. E. Woodward
	\inst{4}
} 

\offprints{F.P. Israel} 
 
\institute{Sterrewacht Leiden, P.O. Box 9513, 2300 RA Leiden, the Netherlands 
\and Joint Astronomy Centre, 660 N. A'ohoku Pl., Hilo, Hawaii, 96720, USA
\and Space Science Applications Laboratory, M2-266, The Aerospace Corporation, 
P.O. Box 92957, Los Angeles, California 90009-2957, USA
\and Department of Astronomy, University of Minnesota, 116 Church Street SE,
Minneapolis, MN 55455, USA}

\date{\dag\,\,Deceased April 4, 2001\\ \\
Received ????; accepted ????} 
 
\abstract{ 
We present maps of the first-ranked HII region complex Hubble~V
in the metal-poor Local Group dwarf galaxy NGC~6822 in the first 
four transitions of $\co$, the 158$\mu$m transition of C$^{+}$, 
the 21-cm line of HI, the Pa$\beta$ line of HII, and the continuum
at 21 cm and 2.2 $\mu$m wavelengths. We have also determined various
integrated intensities, notably of HCO$^{+}$ and near-IR $\h2$ emission.
Although Hubble~X is located in a region of relatively strong HI emission,
our mapping failed to reveal any significant CO emission from it. The 
relatively small CO cloud complex associated with Hubble~V is comparable 
in size to the ionized HII region. The CO clouds are hot ($T_{\rm kin}$ 
= 150 K) and have high molecular gas densities ($n(\h2) \approx 10^{4} 
\cc$). Molecular hydrogen probably extends well beyond the CO boundaries. 
C$^{+}$ column densities are more than an order of magnitude higher than 
those of CO. The total mass of the complex is about 10$^{6}$ M$_{\odot}$ 
and molecular gas account for more than half of this. The complex is 
excited by luminous stars reddened or obscured at visual, but apparent at 
near-infrared wavelengths. The total embedded stellar mass may account for 
about 10$\%$ of the total mass, and the mass of ionized gas for half of 
that. Hubble~V illustrates that modest star formation efficiencies may be 
associated with high CO destruction efficiencies in low-metallicity objects. 
The analysis of the Hubble~V photon-dominated region (PDR) confirms in an 
independent manner the high value of the CO-to-$\h2$ conversion factor 
$X$ found earlier, characteristic of starforming low-metallicity regions.

\keywords{Galaxies: individual: NGC 6822 -- Galaxies: ISM; irregular; 
Local Group -- Radio Lines: ISM; ISM: molecules} 
} 

\authorrunning{F.P. Israel et al.}
\titlerunning{Hubble~V Gas and Dust}
\maketitle

\section{Introduction} 

\begin{table}
\caption[]{(Sub)mm observations log}
\begin{flushleft}
\begin{tabular}{lcccc}
\hline
\noalign{\smallskip}
Transition	& Date    & Freq	& $T_{\rm sys}$ & $\eta _{\rm mb}$ \\
		& (MM/YY) & (GHz) 	& (K)		& 	       \\
\noalign{\smallskip}
\hline
\noalign{\smallskip}
$\co$ (1--0)	& 07/96$^{a}$   & 115		& 425	    	& 0.70  \\
		& 07/97$^{a}$   &		& 435		&	\\
		& 06/98$^{b}$	&		& 370		& 0.73  \\
		& 05/00$^{a}$	&		& 330		& 0.70  \\
$\13co$ (1--0)  & 07/96$^{a}$	& 110		& 235	    	& 0.70 	\\
		& 07/97$^{a}$   &		& 240		&	\\
\noalign{\smallskip}
$\co$ (2--1)    & 06/91$^{c}$   & 230		& 700	  	& 0.69  \\
		& 03/92$^{c}$	&		& 360		&	\\		
		& 06/98$^{b}$	&		& 530		& 0.45  \\
\noalign{\smallskip}
$\co$ (3--2)    & 07/96$^{c}$  	& 345		& 665	  	& 0.58  \\
		& 09/97$^{c}$	&		&		&	\\
$\13co$ (3--2)	& 07/96$^{c}$	& 330   	& 2460    	& 0.58 	\\
\noalign{\smallskip}
$\co$ (4--3)	& 12/94$^{c}$	& 461		& 1280	  	& 0.48 	\\
                & 07/98$^{c}$   &               & 3276          & 0.51  \\
                & 09/98$^{c}$   &               & 1720          &       \\
\noalign{\smallskip}
HCO$^{+}$ (1--0) & 07/96$^{a}$  &  89		& 125	  	& 0.74  \\
		& 07/97$^{a}$	&		& 120		&	\\
\noalign{\smallskip}
CS (3--2)	& 09/96$^{a}$	& 147		& 160	  	& 0.60  \\
\noalign{\smallskip}
H$_{2}$CO$2_{1,2}-1_{1,1}$ & 09/97$^{a}$ & 141	& 174	  	& 0.60  \\
\noalign{\smallskip}
H$_{2}$CO$2_{1,1}-1_{1,0}$ & 09/96$^{a}$ & 150	& 163	  	& 0.60  \\
\noalign{\smallskip}
CI $^{3}P_{1}-^{3}P_{0}$  & 12/94$^{c}$ & 492	& 1765		& 0.43  \\
\noalign{\smallskip}
\hline
\end{tabular}
\end{flushleft}
Notes: $^{a}$ SEST 15 m; $^{b}$ IRAM 30 m; $^{c}$ JCMT 15 m.
\end{table}

NGC~6822 (DDO~209) is a Local Group dwarf irregular galaxy of the 
Magellanic type (IB(s)m), located at a distance of 500 kpc (McAlary 
et al. 1983). Optically, NGC~6822 shows a bar dominated by an irregular 
distribution of OB associations (Wilson 1992a and references therein) 
and HII regions (Hodge et al. 1988). At the northern end of the bar, 
the major HII region complexes Hubble~I, III, V and X (Hubble 1925)
and several relatively luminous OB associations reside in a ridge of 
bright neutral hydrogen. The galaxy is embeddded in a much more extended 
envelope of neutral hydrogen (Brandenburg $\&$ Skillman 1998; de Blok $\&$ 
Walter 2000).

Maps of the infrared emission from NGC 6822, measured with the IRAS 
satellite and processed to a resolution of about 1$'$ (145 pc) were
presented by Israel et al. 1996a (hereafter Paper I). A $J$=1--0 
$^{12}$CO survey of NGC 6822  was the subject of Paper II (Israel 1997a).
In this paper, the third in the series, we present new observations and 
maps primarily of the major star formation complex Hubble~V but also
to some extent of the second major HII region Hubble~X. Coordinates
(epoche 1950.0) are given in Table~2.

The relatively low metal abundance of its HII regions (Hubble~V: [O/H] = 
1.6 $\times$ 10$^{-4}$: Lequeux et al. 1979; Pagel et al. 1980; Skillman
et al. 1989) is consistent with a relatively low time-averaged star 
formation rate (Paper I). This abundance is 
about one third that of the Solar Neighbourhood and right between those 
given by Dufour (1984) for the LMC and the SMC. Also consistent with 
a relatively low star formation rate is the weak radio continuum emission 
from NGC~6822 (Klein et al. 1983). The galaxy has a dust-to-gas ratio of 
about 1.4 $\times$ 10$^{-4}$ (Paper I), which is well within the range 
generally found for dwarf galaxies.

\begin{figure*}[t]
\unitlength1cm
\begin{minipage}[b]{18.0cm}
\resizebox{17.70cm}{!}{\rotatebox{270}{\includegraphics*{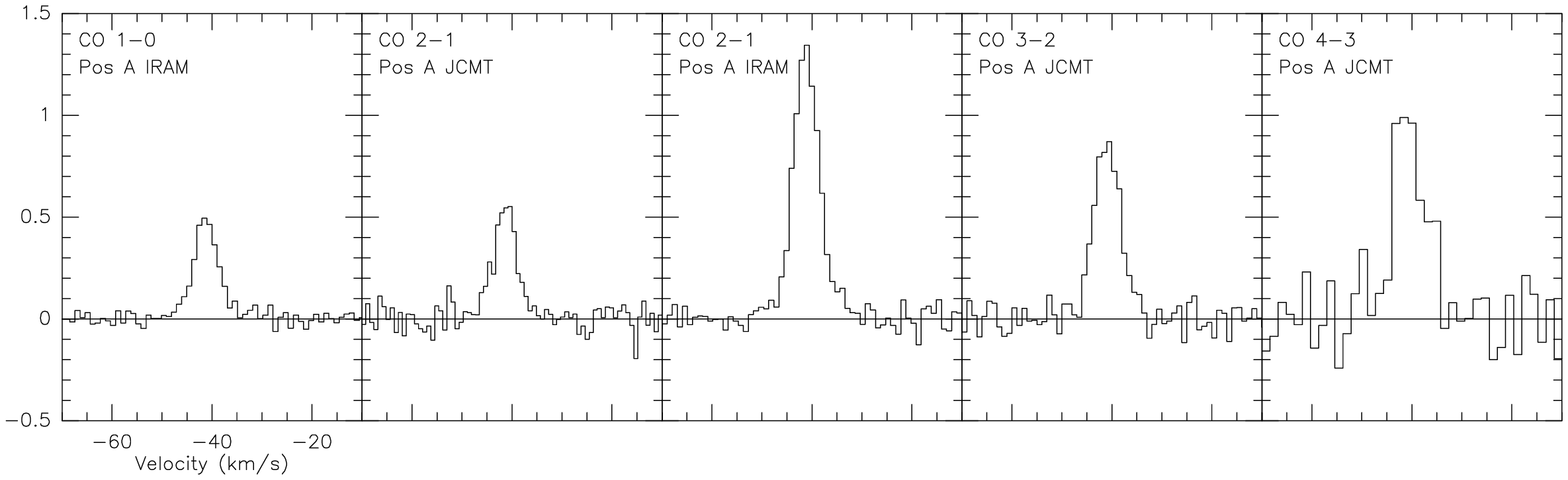}}}
\end{minipage}
\hfill
\begin{minipage}[t]{18.0cm}
\hspace{0.06cm}
\resizebox{17.55cm}{!}{\rotatebox{270}{\includegraphics*{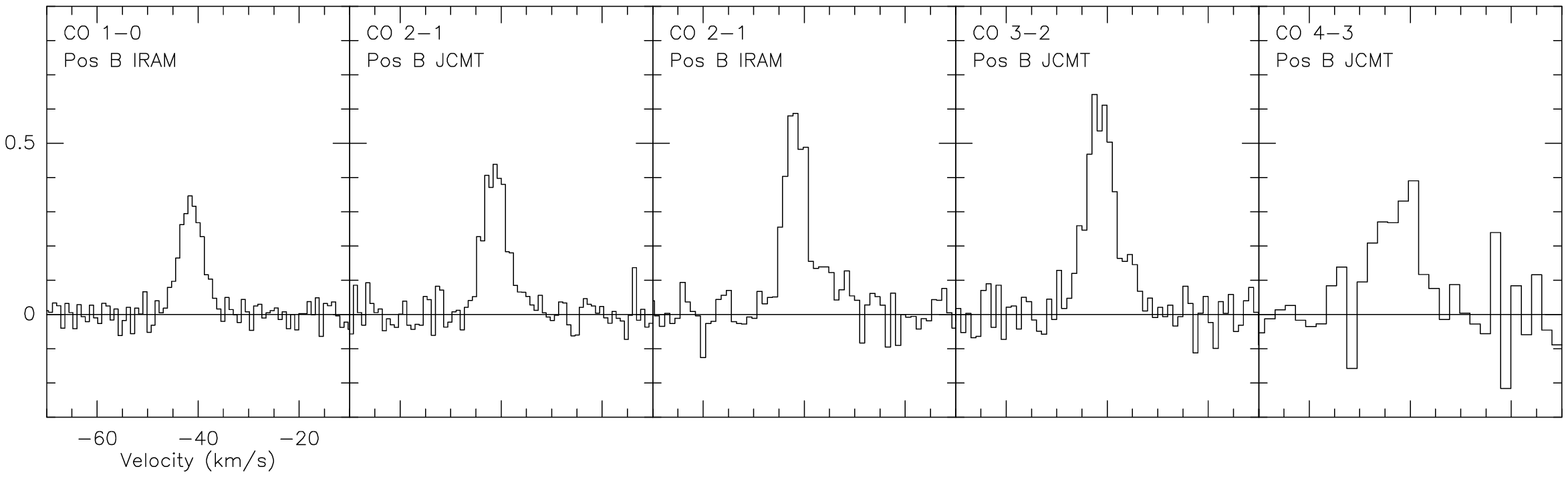}}}
\end{minipage}
\caption{Hubble~V CO profiles of the various transitions. Intensities are 
given in units of $T_{\rm mb} = T_{\rm A}^{*} /\eta_{\rm mb}$; values 
of $\eta_{\rm mb}$ used are those are listed in Table 1. Velocity scale
is $V_{\rm LSR}$.  The spectra in the two leftmost columns were taken
at a resolution of about 22$''$, the spectra in the three rightmost
columns at resolutions of $14'' - 11''$. The spectra at top are those at 
offset position $\Delta \alpha, \Delta \delta$ = 0, +2$''$; the spectra at 
bottom are those at offset position +5$''$, -4$''$; these positions are 
spatially separated by about 8$''$ so that the spectra shown in the two 
rows are not fully independent.}
\label{coprofiles}
\end{figure*}

The brightest HII region complex in NGC~6822 is Hubble~V, connected to
the luminous stars of NGC~6822 OB3 (Wilson 1992a). This association is
about 50 pc in diameter, contains 14 OB stars ($M < 20 M_{\odot}$) with 
an estimated total mass of $850 M_{\odot}$ and has an age of 6.6 million 
years (Wilson 1992b) or less (O'Dell et al. 1999). 
In spite of its optical prominence, Hubble~V has a
radio luminosity only 15$\%$ of that of the first-ranked HII region
complex NGC~604 in M~33; it is comparable to bright Galactic HII regions,
but would not be counted among the very brightest. Nor is it very luminous 
in molecular line emission, although it is the brightest discrete source 
of $J$=1--0 $\co$ emission in NGC~6822 (Paper II). 
Interferometer observations by Wilson (1994) show that about 
half of the single-dish $J$=1--0 $\co$ signal is contributed by compact 
(0.1$'$) components. Wilson (1994) used this result, assuming virial 
equilibrium, to derive a CO-to$\h2$ conversion factor rather similar to 
that of the Solar Neighbourhood, implying relatively small amounts of 
$\h2$. However, Israel (1997a, b) argued that this conversion factor applies 
only to the densest molecular clumps in the complex, and derived for the
entire complex a conversion factor 20 times higher.

\subsection{Molecular line observations}

Details relevant to the observations are listed in Table 1. The system
temperatures given are the means for the respective runs. Most observations
in the 89 - 150 GHz range were obtained with the SEST 15 m telescope at
ESO-La Silla (Chile)\footnote{The Swedish-ESO Submillimetre Telescope (SEST) 
is operated jointly by the European Southern Observatory (ESO) and the 
Swedish Science Research Council (NFR).}. We used the 100/150 GHz SiS 
receivers in dual mode, together with the high resolution acousto-optical 
spectrometer. In split mode, this spectrometer provides two 500 MHz bands 
with a resolution of 43 kHz. At two positions in Hubble~V, $\co$ emission
was measured in the $J$=1--0 and $J$=2--1 transitions with the IRAM 30 m
telescope in Spain.  

\begin{figure*}
\unitlength1cm
\begin{minipage}[b]{18.0cm}
\resizebox{18.0cm}{!}{\rotatebox{270}{\includegraphics*{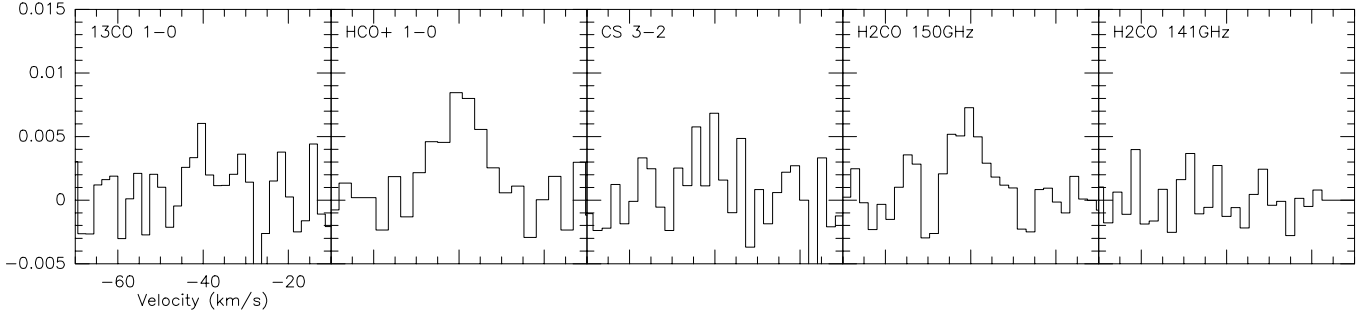}}}
\end{minipage}
\caption[]{Hubble~V. Other lines observed with SEST; intensities are given 
in units of $T_{\rm A}^{*}$ = $\eta_{\rm mb} \times T_{\rm mb}$; values of 
$\eta_{\rm mb}$ are those listed in Table 1. Velocity scale is $V_{\rm LSR}$.
}
\label{molprofiles}
\end{figure*} 

All other observations were made with the JCMT 15 m telescope on Mauna
Kea (Hawaii)\footnote{The James Clerk Maxwell Telescope is operated by the
Observatories on behalf of the Particle Physics and Astrophysics Council
(PPARC) of the United Kigndom, the Netherlands Organization for Scientific 
Research (NWO), and the National Research Council (NRC) of Canada.}.
Up to 1993, we used a 2048 channel AOS backend covering a band of 500 MHz.
After that year, the DAS digital autocorrelator system was used in a band
of 500 and 750 MHz. At 230 GHz, we fully mapped Hubble~V in the $J$=2--1
$\co$ transition; at 345 GHz and 461 GHz, we made small maps of the $J$=3--2
and $J$=4--3 $\co$ transitions covering the emission peak. 
Resulting spectra were binned to various resolutions in order to obtain
the optimum combination of spectral resolution and signal-to-noise ratio.
Usually, only linear baseline corrections were applied to the spectra. 
All spectra were scaled to a main-beam brightness temperature, $T_{\rm mb}$ = 
$T_{\rm A}^{*}$/$\eta _{\rm mb}$; relevant values for $\eta _{\rm mb}$ 
are given in Table 1.

In addition to the measurements of Hubble~V, we also obtained data on
a second and nearby bright HII region complex, Hubble~X, and on a prominent
infrared continuum/millimeter line source at the southern end of the
bar of NGC~6822 (cf. Papers I, II). Even though we mapped the region
containing Hubble~X proper in $J$=1--0 $\co$, we did not detect a signal 
above 30$\%$ of that from Hubble~V. A positive, but noisy result was 
obtained 1$'$ (145 pc) east of Hubble~X. Most likely this 
represents an associated molecular cloud complex.

Spectra are shown in Figs.~\ref{coprofiles} and ~\ref{molprofiles}, 
and summarized in Table 2.
The most remarkable results are (a). the weakness of $\13co$
with respect to $\co$, and (b). the relatively high intensity of
the higher CO transitions with respect to the $J$=1--0 transition.
Both suggest a low optical depth for the $J$=1--0 $\co$ transition.

\begin{figure}
\unitlength1cm
\begin{minipage}[b]{8.7cm}
\resizebox{8.7cm}{!}{\rotatebox{270}{\includegraphics*{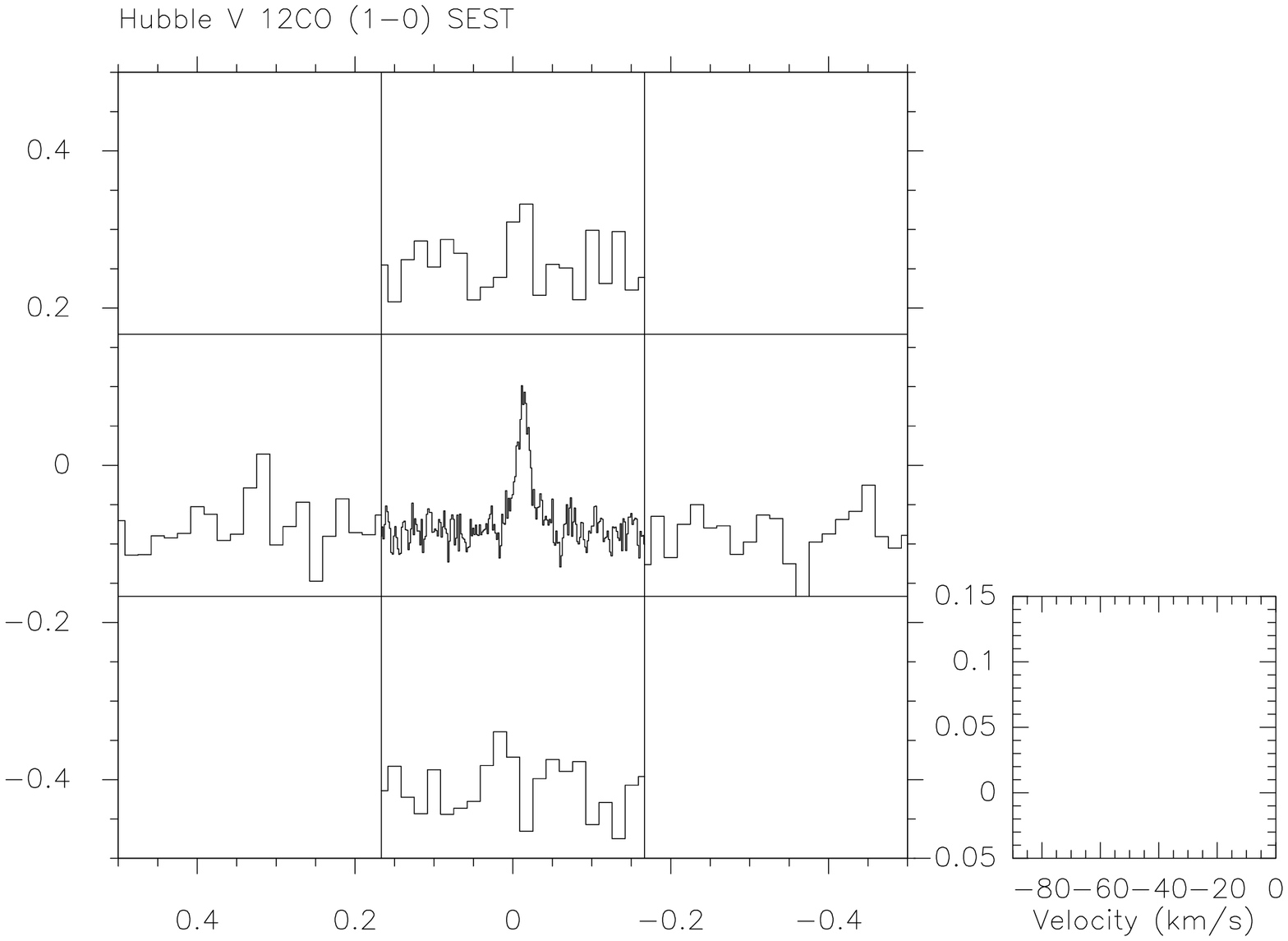}}}
\end{minipage}
\caption[]{$\co$ $J$=1--0 spectra obtained towards Hubble~V at 43$''$ 
resolution. The central profile has a high S/N ratio. The surrounding
spectra at SEST-beam halfpower points have much lower S/N ratios and
therefore are binned to much lower velocity resolutions. Intensities 
are in $T_{\rm A}^{*}$. At 43$''$ (105 pc) resolution, the CO complex
is essentially pointlike. }
\label{cogridmap}
\end{figure}

\begin{table*}
\caption[]{Emission line intenities}
\begin{flushleft}
\begin{tabular}{lccclccccc}
\hline
\noalign{\smallskip}
Source 	& \multicolumn{2}{l}{Position} & Offset & Transition	 & Beam	  & $T_{\rm mb}$ & $\Delta V$ & $V_{\rm LSR}$ & $\int T_{\rm mb}$d$V$ \\
& $\alpha_{\rm o}$(1950.0) & $\delta_{\rm o}$(1950.0)  & $''$   &		 & ($''$) & (mK)	 & $\kms$     & $\kms$	      & $\kkms$  \\
\noalign{\smallskip}
\hline
\noalign{\smallskip}
Hubble~V  & 19:42:03.2 & -14:50:28  & 0, +2  & $\co$ (1--0) 	 & 43 	  &  137$\pm$12	 &  5.6       & -41.3	  & 0.81$\pm$0.06  \\
(Pos. A)  &   	       &   	    &	     &	     	 	 & 23 	  &  489$\pm$31	 &  5.8	      & -41.2	  & 2.70$\pm$0.30 \\
	  &   	       &            &	     & $\13co$ (1--0)    & 45 	  &   10$\pm$7   &  ---	      & -41	  & 0.035$\pm$0.010 \\
	  &   	       &            &	     & $\co$ (2--1) 	 & 21 	  &  545$\pm$26	 &  5.5       & -41.4 	  & 3.18$\pm$0.40 \\
	  &   	       &            &	     &	     	         & 13 	  & 1300$\pm$70	 &  5.6	      & -41.2     & 7.80$\pm$0.90 \\
	  &   	       &            &	     & $\co$ (3--2) 	 & 14 	  &  878$\pm$56  &  6.2       & -41.0     & 5.85$\pm$0.90 \\
	  &   	       &	    & 	     & $\co$ (4--3)	 & 11	  &  949$\pm$59	 &  5.6	      & -41.4	  & 4.74$\pm$0.86 \\
	  &  	       &            &        & HCO$^{+}$ (1--0)  & 57 	  &   14$\pm$3   &  8	      & -39.5     & 0.15$\pm$0.025 \\
	  &  	       &            & 	     & CS (3--2)    	 & 34 	  &    6$\pm$4	 &  ---	      &  ---	  & 0.07$\pm$0.03 \\
& 	  & 	       &	   & H$_{2}$CO $2_{1,2}-1_{1,1}$ & 35 	  &  $<5$        &  ---       &  ---	  & $<0.02$ \\
& 	  & 	       &	   & H$_{2}$CO $2_{1,1}-1_{1,0}$ & 33 	  &   10$\pm$4   &  8         &  40.7     & 0.08$\pm$0.02 \\
\noalign{\smallskip}
Hubble~V  &    	       &	    & +5, -4 & $\co$ (1--0) 	 & 23 	  &  335$\pm$36	 &  5.4	      & -41.4	  & 1.93$\pm$0.21 \\
(Pos. B)  &   	       &	    & 	     & $\co$ (2--1) 	 & 21 	  &  433$\pm$31	 &  5.5	      & -41.3	  & 2.54$\pm$0.30 \\
	  &            &	    &	     &   		 & 13 	  &  588$\pm$81	 &  5.5       & -41.7	  & 3.47$\pm$0.35 \\
	  &   	       &	    &	     & $\co$ (3--2) 	 & 14 	  &  560$\pm$68	 &  6.4	      & -41.4	  & 3.89$\pm$0.55 \\
	  &   	       &	    &	     & $\13co$ (3--2)    & 14 	  & $<150$ 	 &  ---	      &  ---      & $\leq0.6$ \\
	  &   	       &            &	     & $\co$ (4--3) 	 & 11 	  &  354$\pm$96  &  9         & -42.1     & 3.62$\pm$0.90 \\
& 	  &            &  	  & CI $^{3}$P$_{1}-^{3}$P$_{0}$ & 10 	  & $<150$       &  ---       &  ---	  & $<0.9$ \\
\noalign{\smallskip}
\hline
\noalign{\smallskip}
Hubble~X$^{a}$ & 19:42:15.6 & -14:50:31 & 0, 0 & $\co$ (1--0) 	 & 43 	  & $<21$ 	 &  ---       &  ---      & $<0.25$ \\
	  &   	       &            &	     & $\co$ (2--1) 	 & 21 	  & $<70$ 	 &  ---       &  ---      & $<0.27$ \\
\noalign{\smallskip}
	  &   	       &	    & +59, 0 & $\co$ (1--0)      & 43 	  & 60		 &  10	      & -32	  & 0.63$\pm$0.23 \\
	  &    	       &            &	     & $\co$ (2--1) 	 & 21 	  & 50    	 &  8         & -36       & 0.53$\pm$0.15 \\
\noalign{\smallskip}
\hline
\noalign{\smallskip}
South$^{b}$ & 19:41:59.0 & -14:59:54 & 0, 0 & $\co$ (1--0) 	 & 43 	  & 41		 &  21	      &	-50	  & 0.80$\pm$0.14 \\
	  &   	       &            &	     & $\co$ (2--1) 	 & 21 	  & 80	 	 &  18	      &	-44	  & 1.0$\pm$ 0.25 \\
\noalign{\smallskip}
\hline
\end{tabular}
\end{flushleft}
Notes: $^{a}$. Values for Hubble~X represent average result over 3$\times$3 map with 20$''$ spacing.
$^{b}$. Corresponds to IRS 4/CO cloud 5 (Paper I, II); location of NGC~6822 HI maximum.
\end{table*}

\begin{figure*}
\unitlength1cm
\begin{minipage}[b]{5.75cm}
\resizebox{5.95cm}{!}{\includegraphics*{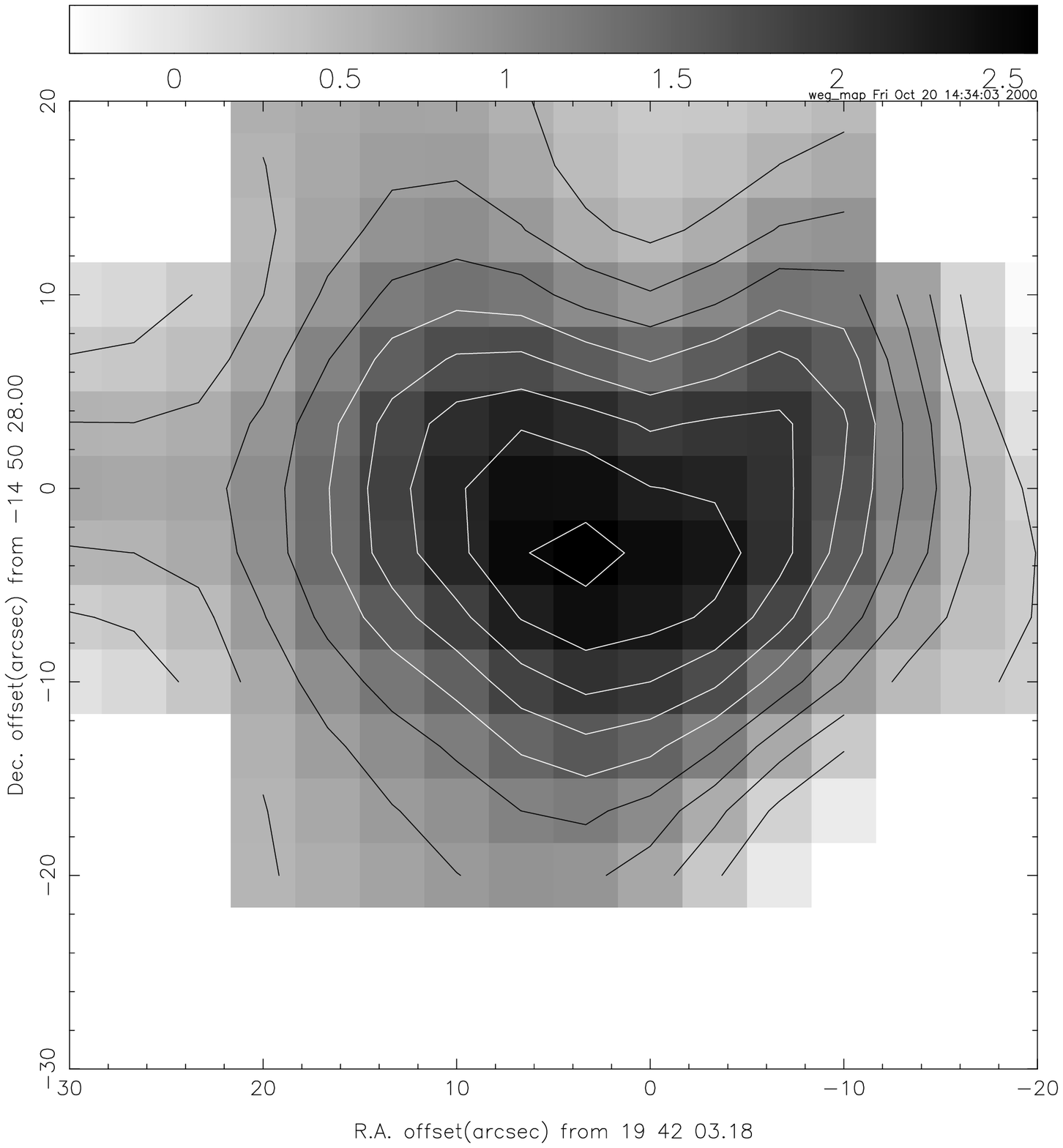}}
\end{minipage}
\hfill
\begin{minipage}[t]{5.75cm}
\resizebox{5.95cm}{!}{\includegraphics*{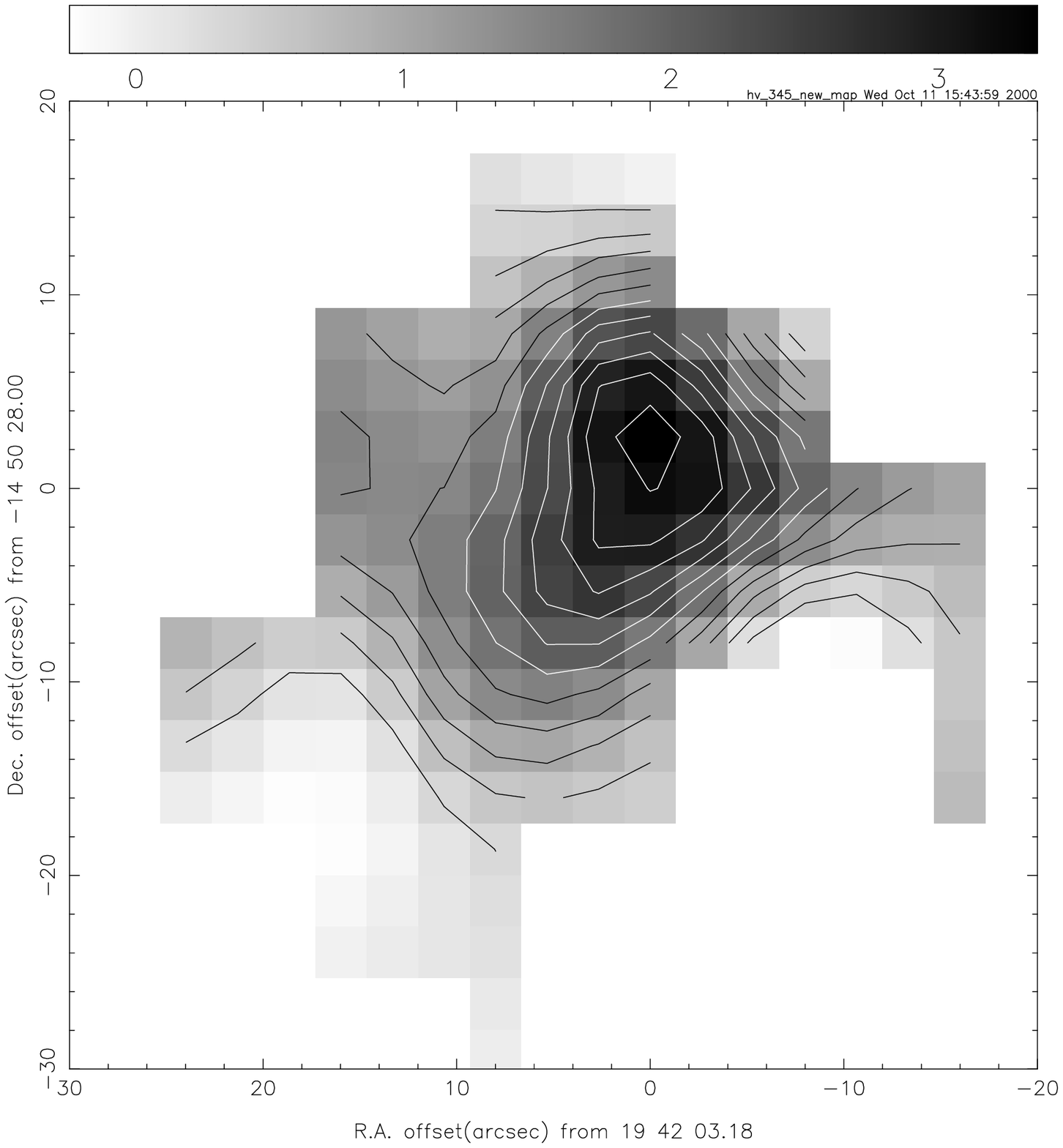}}
\end{minipage}
\hfill
\begin{minipage}[t]{5.75cm}
\resizebox{5.95cm}{!}{\includegraphics*{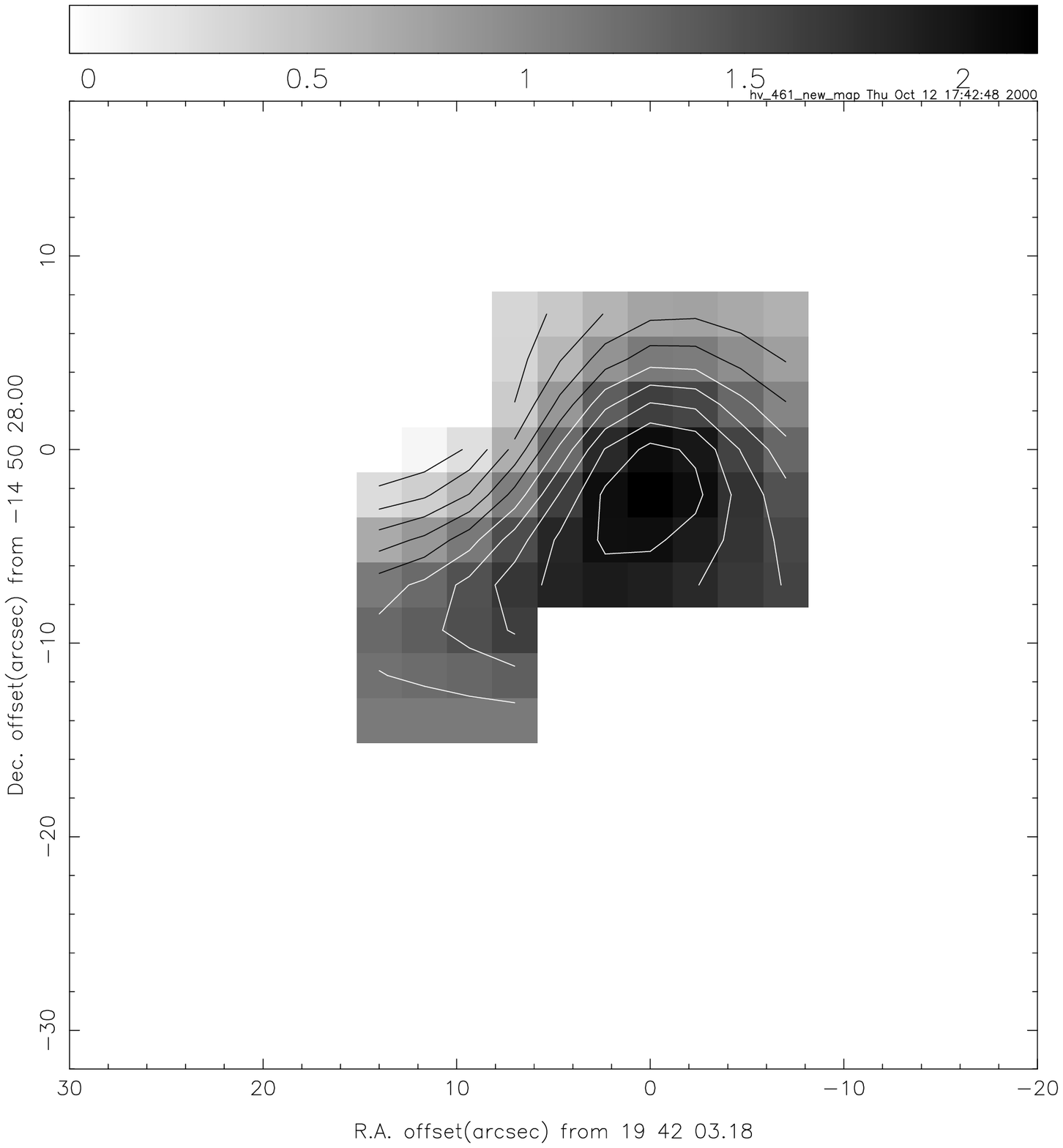}}
\end{minipage}
\caption[]
{Contour maps of velocity-integrated emission from Hubble~V in NGC~6822 
(range: $V_{LSR}$ = -50 to -30 $\kms$) at full resolution. Left to right: 
CO $J$=2--1, CO $J$=3--2 and CO $J$=4--3. Contour values are linear in 
$\int T_{\rm mb} dV$; steps are 0.4 $\kkms$ (2--1), 0.5 $\kkms$ (3--2) 
and 0.4 $\kkms$ (4--3). In all maps, north is at top. Because of
regridding, actual declination of the $J$=4--3 map differs by 2$''$ 
from the printed scale, and is in fact identical to that of the $J$=2--1 
and $J$=3--2 maps.
}
\label{comaps}
\end{figure*}

The spatial extent of emission in the various $\co$ transitions
is shown in Figs.~\ref{cogridmap} and ~\ref{comaps}. The 
five-point $J$=1--0 $\co$ map at resolution 43$''$ (Fig. ~~\ref{cogridmap}) 
shows essentially a point source. The $J$=2--1 CO map 
is peaked at $\Delta \alpha$ = +5$''$, $\Delta \delta$ = -4$''$ with an 
extension to the northwest. The FWHM size of the CO complex is 
$29''\times25''$, which corresponds to a linear size of 50$\times$35 pc 
after correction for finite beamsize. Such a small size is consistent with 
the CO results in the other observed transitions. The $J$=3--2 and $J$=4--3
maps cover most of the emission, but may miss some of the more extended
low-surface brightness emission that appears to be present.

We have determined line ratios in identical beams by convolving the
higher frequency measurements to the lower frequency beamsizes. The results 
are given in Table 5 together with typical values for HII regions in the 
LMC (average of five clouds), the SMC and the starburst core of the
dwarf galaxy He 2-10. We also give the line ratios
for the whole Hubble~V CO source. Because of the limited extent of
our $J$ = 3--2 and especially $J$ = 4--3 map, we may have somewhat 
underestimated the integrated emission in these two transitions, so that 
the ratios given in actual fact could be somewhat higher.

Our single-dish $J$=1--0 and $J$=1--0 CO maps have resolutions of 21$''$ 
and 14$''$ respectively, and may thus be compared to the $J$=1--0 CO 
interferometer map made by Wilson (1994) at a somewhat higher 
resolution of $6''\times 11''$. In her interferometer map, Wilson 
identified three distinct clouds, MC1, MC2 and MC3. The first two
appear to form a single complex connected by a bridge. MC3 appears to be 
a more isolated cloud 30$''$ further north; it is outside our maps. 
The CO emission mapped by us, and shown in Fig. ~\ref{comaps}, corresponds 
to Wilson's cloud complex MC1/MC2, although the detailed resemblance is 
poor. Our $J$=3--2 CO peak is close to MC2, but cloud MC1 and the 
ridge connecting it to MC2 are not easily recognized in Fig.~\ref{comaps}. 
The implications of this are, however, unclear because the 
interferometer map suffers from poor U,V plane coverage and strong 
sidelobe distortion rendering its structural detail uncertain.

\subsection{Atomic line observations}

JCMT observations of the $^{3}P_{1}-^{3}P_{0}$ [CI] transition at 492 GHz,
summarized in Table 2, yielded only an upper limit. In Table 3 we give this 
upper limit expressed in W m$^{-2}$.
The [CII] 158 $\mu$m observations were made with the MPE/UCB Far-Infrared 
Imaging Fabry-Perot Interferometer (FIFI; Poglitsch et al. 1991) on the
NASA Kuiper Airborne Observatory (KAO) in April 1992. FIFI had a 5$\times$5
focal plane array with detectors spaced by 40$''$ (Stacey et al. 1992). Each 
detector had a FWHM of 55$''$ and the beam shape was approximately Gaussian 
(68$''$ equivalent disk; beam solid angle $\Omega_{\rm B}$ = 8.3 $\times$
10$^{-8}$ sr). We observed in `stare mode' (velocity resolution 50 $\kms$)
by setting the bandpass of the Fabry-Perot to the line center at the 
object velocity. Observations were chopped at 23 Hz and beam-switched to two
reference positions about 6$'$ away. The data were calibrated by observing an
internal blackbody source. The calibration uncertainty is of the order of 
30$\%$ and the absolute pointing uncertainty of the array is better than
15$''$. 

\begin{figure}
\unitlength1cm
\begin{minipage}[]{8.75cm}
\resizebox{8.5cm}{!}{\rotatebox{270}{\includegraphics*{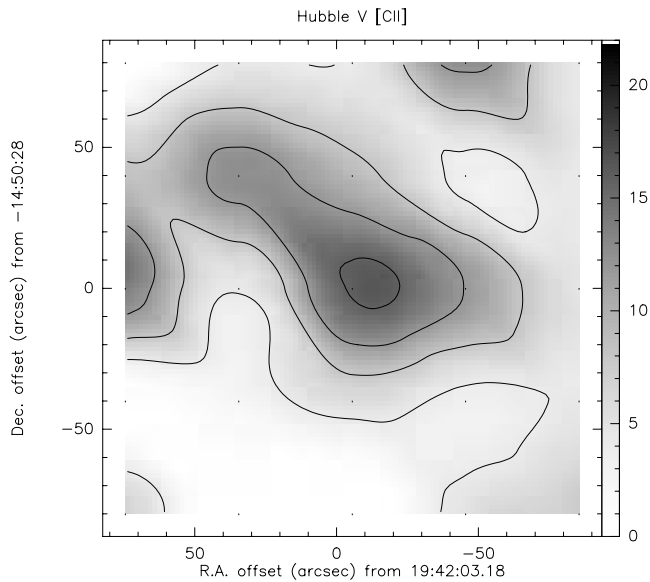}}}
\end{minipage}
\caption[]{Map of [CII] emission centered on Hubble~V. Contours are 
linear with both first level and interval of $4 \times 10^{-9}$ W
m$^{-2}$ sr$^{-1}$. 
}
\label{ciimap}
\end{figure}

The resulting map is shown in Fig.~\ref{ciimap}. The [CII] emission is 
relatively weak. There is a clear peak at 
the position of Hubble~V, and a second peak at $\Delta \alpha$ = -40$''$, 
$\Delta \delta$ = +80$''$, i.e. 215 pc northwest of Hubble~V. A large part 
of the field is filled by weak and diffuse [CII] emission with a surface 
brightness of approximately 8 $\times$ 10$^{-9}$ W m$^{-2}$ sr$^{-1}$. The 
[CII] emission directly associated with Hubble~V has a larger extent than 
the CO emitting region, but the poor [CII] resolution makes it impossible 
to be more quantitative. 

\begin{figure}
\unitlength1cm
\begin{minipage}[]{8.75cm}
\resizebox{8.5cm}{!}{\rotatebox{270}{\includegraphics*{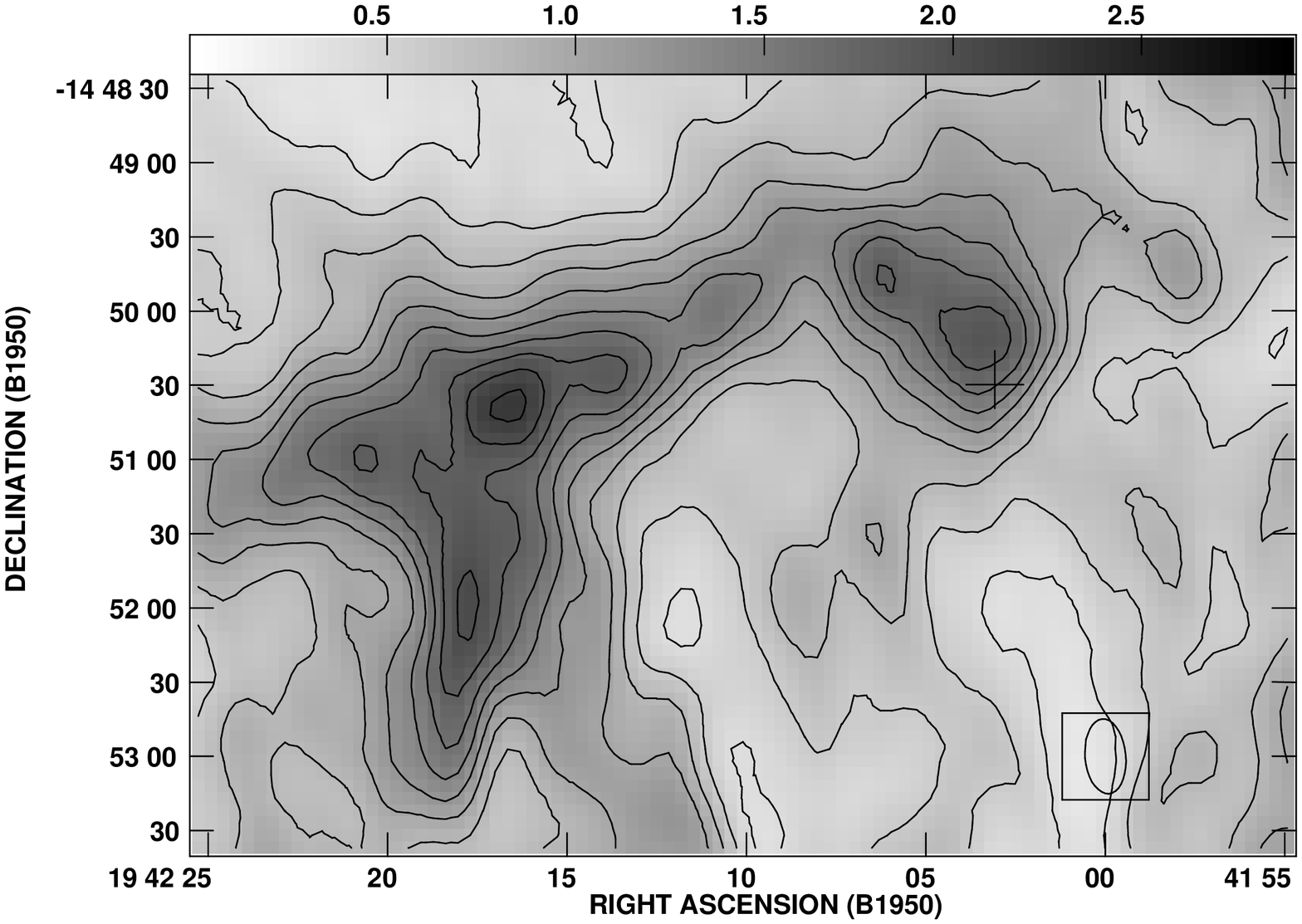}}}
\end{minipage}
\begin{minipage}[]{8.75cm}
\resizebox{8.5cm}{!}{\rotatebox{270}{\includegraphics*{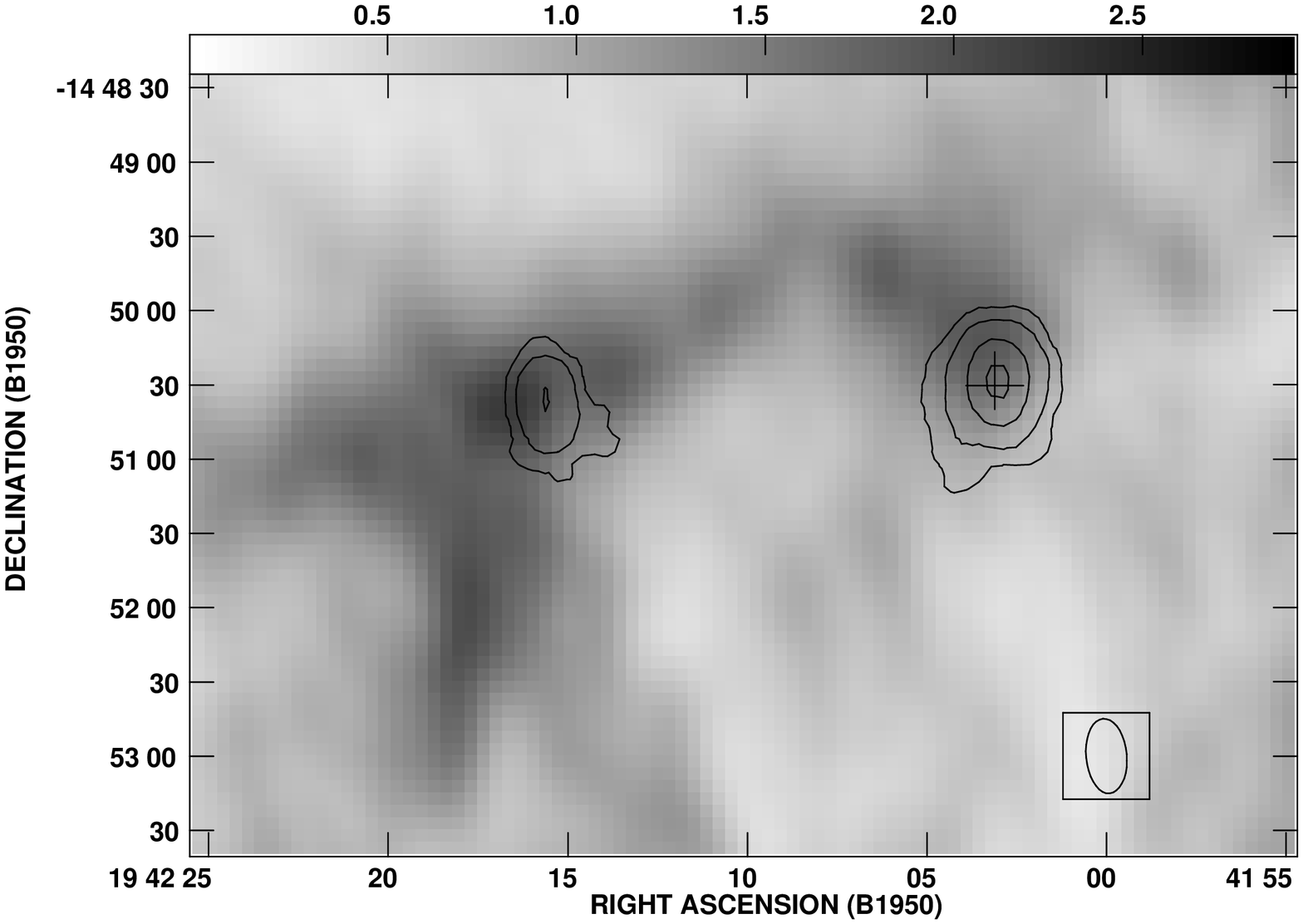}}}
\end{minipage}
\caption[]{Comparison of HI 21-cm total column density and 21-cm radio 
continuum VLA observations of NGC 6822 in the vicinity of Hubble~V and 
Hubble~X. Top: map of total HI column density in both greyscale and 
contour representation. The greyscale is labeled in units of 10$^{21}$ 
H atoms cm$^{-2}$ with contour levels from 0.2 to 2.2 ascending in 
increments of 0.2. Bottom: map of radio continuum emission at 21 cm
in contour representation plotted on top of the HI column density in 
greyscale representation (identical to top). The 21-cm radio continuum 
contours are plotted at values of 1, 2, 5, and 10 mJy/Beam.  In both 
figures, a cross marks the (0,0) position of the CO and [CII] maps; the
size of the cross corresponds to $J$=2--1 CO beamsize. The HI synthesized 
beamsize of 30.3$\arcsec$ $\times$ 16.4$\arcsec$ is shown by the ellipse 
in the box in the lower right hand corner.
}
\label{himaps}
\end{figure}

The HI observations of NGC~6822 with the Very Large Array\footnote{The
Very Large Array is a facility of the National Radio Astronomy Observatory}
consisting of 8 hours (essentially a full transit) in the ``D'' 
configuration on 7 September 1992 and another 8 hours in the ``CnB'' 
configuration of 5 June 1993 under program AW312 (PI: C.\ Wilson).  
The ``CnB'' configuration was chosen for the higher resolution 
observations because the galaxy lies at a declination of $-14^{\circ}$, 
resulting in a strongly elliptical synthesized beam for normal 
configurations.  The ``D'' configuration was chosen for the lower 
resolution observations in order to maximize the observing time with 
the Sun below the horizon. The observations were taken at a single 
pointing (RA $=19^{h}42^{m}07^{s}$, Dec.\ $= -14^{\circ}55^{'}42^{''}$, 
1950.0), with the correlator in ``2AD'' mode consisting of 127 channels 
with widths of 2.5 km s$^{-1}$ (after on-line Hanning smoothing)
centered on a heliocentric velocity of $-$55 km s$^{-1}$. Total on-source 
integration time was 383 minutes for the D configuration observations
and 340 minutes for the CnB configuration observations. The images were 
reconstructed with robust weighting resulting in a synthesized beam of 
$30.3\arcsec \times 16.4\arcsec$. Preliminary results were reported in 
Brandenburg $\&$ Skillman (1998), and a more complete report is planned.

The HI distribution shows a local maximum in the atomic hydrogen 
distribution just north of Hubble~V. Emission then extends westwards in a 
clumpy ridge towards Hubble~X. This nebula is likewise located at the 
edge of a local atomic hydrogen maximum, which is in fact somewhat
stronger than the Hubble~V maximum. Although the [CII] and HI distributions
are not identical, comparison of Figs.~\ref{ciimap} and ~\ref{himaps} 
shows that they follow one another, at least over the region mapped in
[CII]. The two HII regions, as traced by their 21-cm continuum emission,
are both located close to, but clearly offset from, the ridge of maximum HI
intensity. Note that the relative positions of HII and HI are extremely
accurate, as they are both derived from the same dataset.

As is the
case for HI, [CII] emission likewise extends in a northeast-southwest ridge 
from Hubble~V to Hubble~X. In Table 3, we have listed the observed
[CII] intensities for the pixel containing Hubble~V (Source Peak), for the 
extended source of which this pixel is part (Source Total), and for the 
whole observed field (Field Total).

\subsection{Near-infrared hydrogen line and continuum observations}

We obtained a near-infrared spectrum, covering the wavelength region 
from 2.09 to 2.21 $\mu$m, in April 1987 with the UK Infrared Telescope 
(UKIRT), using a single-channel InSb detector equipped with a circular 
variable filter wheel of constant spectral resolution $\lambda/\Delta 
\lambda$ = 120. The aperture was 19.6 $''$, covering most of the complex. 
Intensities were calibrated and corrected for atmospheric transmission 
by observations of the standard stars BS~3903, BS~4550 and BS~6220 ($K$ 
= 2.04, 4.39 and 1.39 mag, and $T_{\rm eff}$ = 4800, 5200 and 4700 K 
respectively) observed at similar airmass. The $v$=1--0 S(1) $\h2$ and 
Brackett-$\gamma$ lines are clearly detected with equivalent widths of 
11 and 34 nm respectively. Line intensities are given in Table 3.

We also used the near-infrared Fabry-Perot imaging spectrometer (FAST;
Krabbe et al. 1993) at the Cassegrain focus of the 4.2 m William Herschel 
Telescope (WHT) at La Palma, Spain, to obtain images of Hubble~V in the 
Br-$\gamma$ and $v$=1--0 S(1) lines. The FAST camera used an SBRC
58$\times$52 InSb array with a pixel scale of 0.5$''$ and a field of
about $30''\times 30''$. A spectral resolution of $\lambda/\Delta
\lambda$ = 950 was provided by a scanning Fabry-Perot interferometer with
a circular variable filter wheel as order-sorter. Images were obtained at
the Hubble~V central velocity, and at velocities offset from the line 
by about 300 $\kms$. After subtraction of the dark current, the images 
were flat-fielded and sky-subtracted. The resulting line-plus-continuum 
images were corrected for atmospheric transmission and instrumental 
response with the use of the standard stars BS~3888 and BS~4550. As a
last step, the mean of the continuum on either side of the line was 
subtracted from the image containing the line.

Hubble~V was detected in Brackett-$\gamma$ as an amorphous structure of
4$''\times$5$''$ FWHM. The integrated flux determined from the map is 
1.0 $\times$ 10$^{-16}$ W m$^{-2}$, uncertain by a factor of two. As the 
nominal flux is very close to that observed in the 19.6$''$ aperture of 
the spectroscopic observations, we take this agreement to indicate that 
the value in Table 3 is close to the total Brackett-$\gamma$ line 
intensity of Hubble~V. The $\h2$ image did not yield a detection. The 
r.m.s. noise in the map is 2 $\times$ 10$^{-8}$ W m$^{-2}$ sr$^{-1}$. 
The two-$\sigma$ upper limit to the surface brightness and the line flux 
given in Table 3 then suggest that the $\h2$ emission is relatively smooth 
and extended on a scale larger than 6$''$. For comparison purposes, we 
have also listed the H$\alpha$ flux integrated over the whole HII region 
complex  determined by Hodge et al. (1989). According to Fig.~1 of Hodge 
et al. (1989), our 19.6$''$ (48 pc) aperture should contain about 70$\%$ 
of the total Brackett-$\gamma$ flux.  

Finally, Hubble~V was observed in August 1998, in J, H and K$_{\rm s}$
broadband filters, as well as in narrow-band filters centered on the
wavelengths of the Pa$\beta$ and (1--0) S(1) $\h2$ lines. The K$_{\rm s}$
bandpass is an abbreviated version of the standard $K$-band filter, spanning
the wavelength range from 1.98 to 2.32 $\mu$m, chosen to minimize the
thermal background. The near-infrared images were acquired with the 
Aerospace Corp. near-infrared camera. This is a liquid nitrogen cooled imager 
based on the NICMOS3 detector.  When configured for use at the Wyoming 
Infrared Observatory the camera has a 110 x 110 arcsecond field of view.  
A more detailed description is provided by Rudy et al. (1997). 
The standard star employed for the August 1998 images was HD 225023 
whose near-infrared magnitudes are tabulated by Elias (1982). The 
Pa$\beta$ images were acquired using a narrow passband filter (FWHM = 122 
angstroms); the Pa$\beta$ flux was calculated by using the J-band image
to estimate and remove the continuum emission. The full extent of Pa$\beta$
emission is about 6$''$, and the nebula shows little structural detail.

\begin{table}
\caption[]{Hubble~V. Carbon and hydrogen line emission}
\begin{flushleft}
\begin{tabular}{lccr}
\hline
\noalign{\smallskip}
Transition			 & Resol.  & $F_{\rm line}$    & Remarks \\
				 & ($''$)  & (W m$^{-2}$)      & \\
\noalign{\smallskip}
\hline
\noalign{\smallskip}
CI $^{3}$P$_{1}-^{3}$P$_{0}$ 	 & 10 	   & 	  $<$7.4 (-19) & \\
\noalign{\smallskip}
CII $J=\threehalf - \half ^{2}$P & 55      & 1.6$\pm$0.2 (-15) & Source Peak \\
		        	 &	   & 2.8$\pm$0.4 (-15) & Source Total \\
		        	 &	   & 6.9$\pm$0.9 (-15) & Field Total \\
\noalign{\smallskip}
$\h2$ $v$=1--0 S(1)		 & 19.6	   & 2.8$\pm$0.7 (-17) & \\
\noalign{\smallskip}
HII Br$\gamma^{a}$   		 & 19.6	   & 8.4$\pm$1.0 (-17) & \\
\noalign{\smallskip}
HII Pa$\beta$   		 & ---	   & 4.2$\pm$1.0 (-16) & Bright Only \\
\noalign{\smallskip}
HII H$\alpha^{b}$	         & ---	   & 4.8$\pm$0.5 (-15) & Source Total \\
\noalign{\smallskip} 
FIR		                 & ---     & 4.7$\pm$1.2 (-13) & Paper I \\ 
\noalign{\smallskip}
\hline
\end{tabular}
\end{flushleft}
Notes: $^{a}$. Corresponding to about 70$\%$ of source total (see text)
$^{b}$. O'Dell et al. (1999)
\end{table}

\begin{table}
\caption[]{Hubble~V. Near-infrared photometry}
\begin{flushleft}
\begin{tabular}{lcccl}
\hline
\noalign{\smallskip}
Object	 & $K_{\rm s}$  & $H-K_{\rm s}$  & $J-H$ & Type \\
	 & \multicolumn{3}{c}{(mag)} \\
\noalign{\smallskip}
\hline
\noalign{\smallskip}
North    & 14.52   & +0.06  & +0.55 & Foreground Star \\
\noalign{\smallskip}
Center   & 15.35   & +0.16  & +0.67 & Visible Cluster \\
\noalign{\smallskip}
South    & 15.93   & $\geq$ 1.07 & --- & Obscured Cluster \\
\noalign{\smallskip}
\hline
\end{tabular}
\end{flushleft}
\end{table}

\section{Analysis}

\subsection{Hubble~V foreground extinction}

Assuming intrinsic line ratios H$\alpha$/Pa$\beta$ = 17.4 and 
H$\alpha$/Br$\gamma$ = 104.4 (Pengelly, 1964; Giles 1977), our line
fluxes (Table~3) predict the source-averaged reddening to be
$ 0.35 < E(B-V) < 0.55$ mag with a total deredenned flux 
S(H$\alpha) = 1.2\pm0.2 \times 10^{-14}$ W m$^{-2}$. This reddening is 
higher than the (mostly Galactic foreground) reddening $E(B-V)$ = 
0.24 -- 0.36 mag (Gallart et al. 1996; Massey et al. 1995; McAlary et al. 
1985). Although our value is comparable to the median reddening of early 
type stars in NGC~6822 ($E(B-V)$ = 0.45 mag (Wilson 1992b; Bianchi et al. 
2001), it must be considered a lower limit, because neither the
Pa$\beta$ nor the Br$\gamma$ fluxes represent all the flux of 
Hubble~V sampled in H$\alpha$. 

With flux-densities $S_{\rm 1.5GHz}$ = 20 mJy, $S_{\rm 4.8GHz}$ = 17 
mJy and $S_{\rm 10.7GHz}$ = 16 mJy (Condon 1987; Klein $\&$ Gr\"ave 
1986; Klein et al. 1983) the radio continuum emission from Hubble~V 
(Fig.~\ref{himaps}) is clearly thermal and optically thin. From these 
radio continuum flux-densities and the H$\alpha$ flux we obtain for $T_{e}$ = 
11 500 K (Lequeux et al. 1979; Skillman et al. 1989) a more accurate
{\it overall} reddening $E(B-V)$ = 0.65 mag. Thus, the nebula 
appears to suffer a higher mean reddening than NGC~6822 as a
whole. This agrees, at least qualitatively, with the finding that
in the Magellanic Clouds the younger stellar population suffers
significantly more reddening than the older population (Zaritsky 1999;
Zaritsky et al. 2002). However, in the next section we will show
that in actual fact much of the nebula is only modestly reddened, whereas 
a part is almost completely obscured at visual wavelengths, and becomes
progressively more visible at near-infrared wavelengths only.

\begin{table*}
\caption[]{Hubble~V. Normalized line intensities}
\begin{flushleft}
\begin{tabular}{lcccccccc}
\hline
\noalign{\smallskip}
Transition	& Resol. & \multicolumn{7}{c}{Normalized Intensity} \\
		&	 & \multicolumn{3}{c}{Hubble~V}			& \multicolumn{2}{c}{LMC$^{a}$} & SMC$^{b}$ & He 2-10$^{c}$ \\
		& ($''$) & Pos. A. 	 & Pos. B 	& Total 	  & 30 Dor & Other & LIRS 36 & \\
\noalign{\smallskip}
\hline
\noalign{\smallskip}
$\co$ (1-0)	& 43	 &  ---	   	 & ---        	& 0.6$\pm$0.1	  & 0.95   & 0.9   & 0.8     & 1.0  \\
		& 21	 &  0.9$\pm$0.1	 & 0.8$\pm$0.1	& ---          	  & ---	   & ---   & ---     & ---  \\
$\co$ (2-1)     & --	 &  1	   	 & 1       	& 1          	  & 1      & 1     & 1       & 1    \\
$\13co$ (2-1)   & 21     &  ---		 & ---		& ---		  & 0.15   & 0.14  & 0.13    & 0.05 \\
$\co$ (3-2)     & 14	 &  0.9$\pm$0.2  & 1.1$\pm$0.1	& 1.2$\pm$0.2 	  & 1.6    & 0.9   & 0.9     & 1.3  \\
$\co$ (4-3)	& 14	 &  0.5$\pm$0.1  & 1.0$\pm$0.1	& 0.7$\pm$0.1 	  & ---    & ---   & ---     & ---  \\
$\13co$ (3-2)	& 14	 &  ---		 & $\leq$0.16	& ---		  & ---    & ---   & ---     & 0.08 \\
CI $^{3}$P$_{1}-^{3}$P$_{0}$ & 14 & ---	 & $<$0.25  	& ---		  & ---    & ---   & ---     & ---  \\
\noalign{\smallskip}
CII $J=\threehalf - \half ^{2}$P  & 55  & --- & ---	& 13$\pm$5  	  & 25 	   & 5     & 10      & ---  \\
\noalign{\smallskip}
\hline
\noalign{\smallskip}
$\co$ (1-0)	& --	 &		 &		& 1		  & 1	   & 1     & 1       & 1    \\
$\13co$ (1-0)   & 43	 &  ---		 &		& 0.043$\pm$0.012 & 0.09   & 0.10  & 0.11 & $<$0.05 \\
HCO$^{+}$ (1-0) & 57 	 &  ---		 &		& 0.31$\pm$0.06   & 0.29   & 0.14  & 0.07    & ---  \\
CS (3-2)	& 34	 &  ---		 &		& 0.06$\pm$0.03	  & 0.03   & 0.03  & 0.02    & ---  \\
H$_{2}$CO	& 35	 &  ---		 &		& 0.07$\pm$0.03	  & 0.04   & 0.03  & 0.02    & ---  \\
\noalign{\smallskip} 
\hline
\end{tabular}
\end{flushleft}
Notes: a. LMC data: Johansson et al. (1994); Israel et al. (1996a); 
Chin et al. (1997); Heikkil\"a et al. (1999). b. SMC data: Chin et al. 
(1998); Israel (unpublished); c. Baas et al. 1994.
\end{table*}

\subsection{The Hubble~V radiation field}

\begin{figure*}
\unitlength1cm
\begin{minipage}[b]{17.5cm}
\resizebox{17.0cm}{!}{\includegraphics*{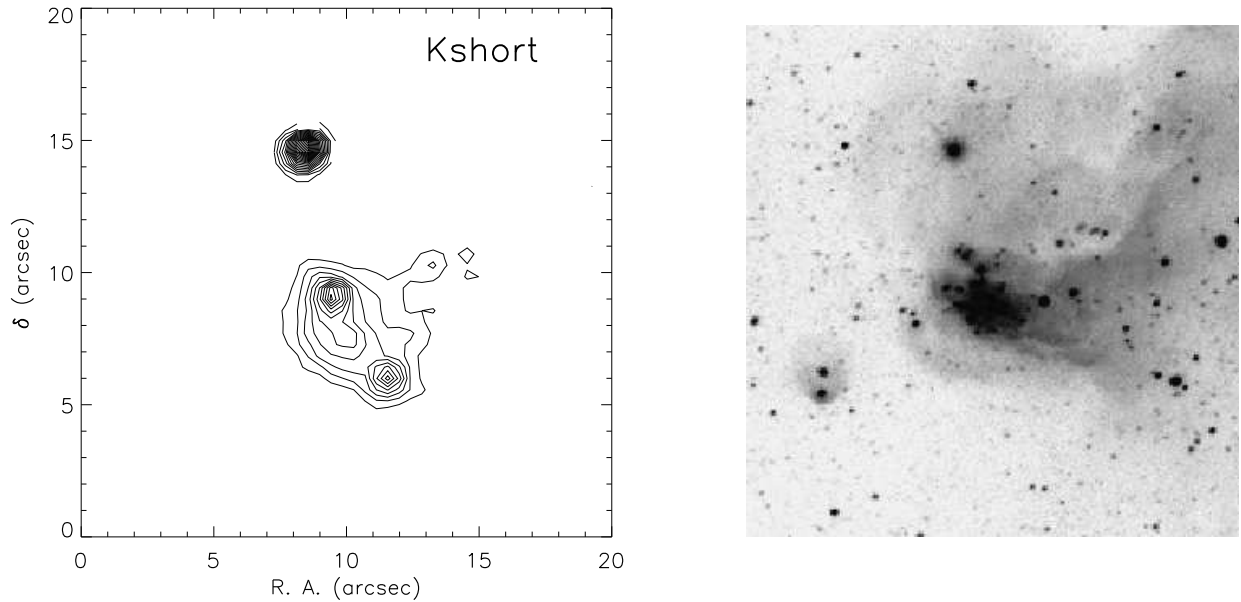}}
\end{minipage}
\caption[]{Left: Near-infrared K-band image of Hubble V and its 
surroundings. As no absolute position is available, only offsets
relative to a randomly chosen position are supplied. 
Right: Corresponding part of the NASA/ESA/STScI 
HST image based on data from O'Dell et al (1999) and Bianchi et 
al. (2001). The northern Galactic foreground star and the stellar 
cluster in Hubble~V, easily seen in both images, were used to bring
the HST image and the K-band image to the same scale and line-up.
The southern $K_{\rm s}$ band 
object, presumably representing an embedded cluster, has no visual 
counterpart. Most of the extended nebular emission apparent in the 
near-IR image is likewise heavily obscured. Full-color HST image 
may be found on the web at http://heritage.stsci.edu/2001/39/
}
\label{optical&kmap}
\end{figure*}

The H$\alpha$ map by Collier $\&$ Hodge (1994) exhibits a typical 
core-halo structure. More detail is supplied by an HST image of 
Hubble~V, based on data procured and discussed by O'Dell et al. 
(1999) and Bianchi et al. (2001). This image (Fig.~\ref{optical&kmap})
shows the core to contain a cluster of luminous stars, and the halo to 
consist of diffuse and filamentary structues against which several
OB association member stars can be seen (cf. Bianchi et al. 2001). 
Most relevant to us are the clear indications of substantial 
obscuration in the northeastern and especially southeastern part of 
the complex. The core appears to be the region directly in contact
with the molecular cloud complex, whereas the halo is mostly ionized 
gas expanding into surrounding space (cf. O'Dell et al, 1999). The 
individually identified early type stars (see e.g. Massey et al. 
1995) are all in the relatively diffuse western halo part of the 
nebula. The excitation parameter of an HII region complex is defined 
as $u = 14.2 (S_{4.8} D^{2} T_{e}^{0.35})^{1/3}$ with flux-density
$S_{\nu}$ in Jy, distance $D$ in kpc, and $T_{e}$ in 10$^{4}$ K
(cf. Mezger $\&$ Henderson 1967).
Using the radio flux-densities from the previous section, we find
for the Hubble~V HII region complex $u$ = 234 pc cm$^{-2}$. This
implies a {\it minimum} Lyman continuum photon flux $N_{\rm L}$ = 
8.45 $\times$ 10$^{50}$ photons s$^{-1}$ (Panagia 1973), corresponding 
to the UV output of more than a dozen O5 stars. From the compilation 
by Wilson (1992a) we may estimate that previously identified 
early-type stars can contribute, at face value, no more than about 
three quarters of this {\it minimum} required Lyman continuum flux. 
Moreover, as the HII region appears to be partly density-bounded, the 
actually required flux should be higher by mpore than a factor of two, 
thus relegating the role of the identified stars to that of minor 
contributors. Moreover, these estimates assume that the identified stars 
are {\it embedded} in the nebula. As the HST image shows that many of 
these stars may be detached from the nebula, this is {\it not} the case,
and their contribution to the {\it actual} excitation of Hubble~V is 
even less, and probably quite minor if not negligible.

Our $K_{\rm s}$-band image of Hubble~V and its surroundings 
(Fig.~\ref{optical&kmap}) shows three unresolved objects, as well as nebular 
emission (primarily Br$\gamma$) extending into the southeastern obscuration. 
The HST image and the infrared measurements summarized in Table
4 suggest that the northernmost object is a relatively unreddened star 
of late spectral type in the Milky Way foreground. The central object 
($K_{\rm s}$ = 15.35 mag) coincides with a cluster of stars in the HST 
image (Fig.~\ref{optical&kmap}). The infrared colours suggest 
that the emission is dominated by light from K (super)giants, again 
with relatively little reddening. However, the southernmost object 
has no optical counterpart and instead coincides with a region of 
high obscuration in the HST image. It is not even seen in the 
$J$ and $H$ bands ($H \geq$ 17.0 mag and $J \geq$ 17.6 mag),
indicating an extinction at $K$ of the order of 1.5 mag, i.e. $A_{\rm V} 
\geq$ 17 mag ($E(B-V) \geq$ 5.4 mag). Most likely, it represents a 
cluster of luminous stars responsible for a significant part, if 
not all, of the excitation of Hubble~V and the heating of the 
molecular cloud observed by us.

\begin{figure}
\unitlength1cm
\begin{minipage}[b]{8.7cm}
\resizebox{8.7cm}{!}{\includegraphics*{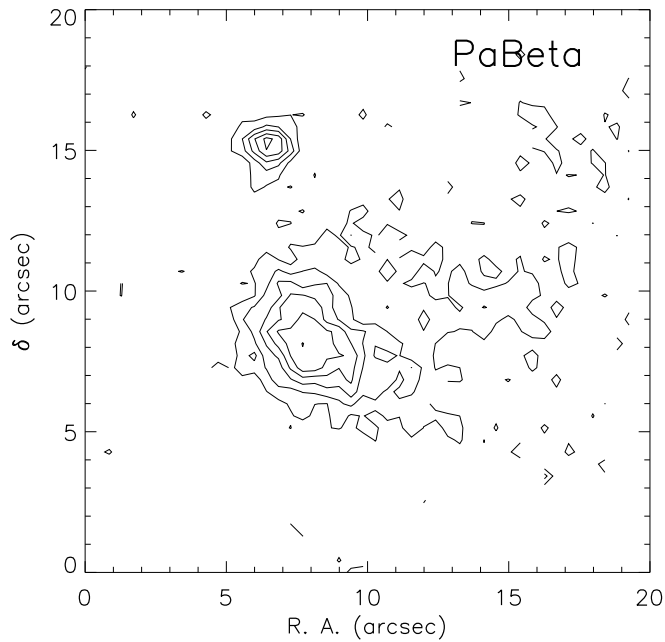}}
\end{minipage}
\caption[]{Near-infrared Paschen-$\beta$ line image of Hubble V and its 
surroundings. Again only offsets relative to a randomly chosen position 
are supplied. Contours are linear in multiples of the lowest contour; 
contour step is 16.98 mag per square arcsecond corresponding to 
$4.5 \times 10^{-18}$ W m$^{-2}$ arcsec$^{-2}$.
The bright core, presumably containing ionization fronts, 
is well depicted; the more diffuse extended emission to the north and west
is more poorly represented. The unresolved object north of the nebula
is a residual continuum emission from an imperfectly subtracted bright
foreground star - see Fig.~\ref{optical&kmap}. 
}
\label{pabetamap}
\end{figure}

In (Fig.~\ref{pabetamap}) we show the nebula imaged in the Pa$\beta$ 
line. Comparison of the distribution of line emission with respect to
the (northern) foreground star show that it is brightest between the 
central and southern clusters, i.e. likewise in a region 
obscured at visual wavelengths. The nebular complex radius is about
20 pc. Taking the HII region-derived Lyman flux and assuming this to 
originate in a point source separated by 20 pc from the HII 
region/neutral cloud interface represented by the bright core, 
we find the strength of the UV radiation 
field at the interface to be $I_{\rm UV}$ = 725. Here, the unit chosen 
is such that $I_{\rm UV}$ = 1 corresponds to a flux $I_{1000} = 4.5 
\times 10^{-8}$ photons s$^{-1}$ cm$^{-2}$ Hz$^{-1}$ (Black $\&$ van 
Dishoeck 1987). Alternatively, we may express the strength of the 
radiation field as $G_{\rm o} \approx$ 300, in units of 12.6 $\times 
10^{-6}$ W m$^{-2}$. This makes Hubble~V similar to the LMC HII 
regions N~159 and N~160 (see Israel et al. 1996b; Bolatto et al. 2000).

\subsection{The Hubble~V PDR}

\begin{table*}
\caption[]{Hubble~V. Model parameters}
\begin{flushleft}
\begin{tabular}{lcccccccccccccc}
\hline
\noalign{\smallskip}
Pos. & \multicolumn{4}{c}{Component 1}          & \multicolumn{4}{c}{Component 2}                         & \multicolumn{6}{c}{Line Ratios$^{a}$} \\
 	 & $T_{\rm k}$ & $n(\h2)$ & $\nco$ & Weight & $T_{\rm k}$ & $n(\h2)$ & $\nco$  & Weight & $R_{12}$ & $R_{32}$ & $R_{42}$ &  $R_{1}$ & $R_{2}$ & $R_{3}$ \\
         & (K)	&  ($\cc$)  & ($\cmk$) &        & (K)    &  ($\cc$) & ($\cmk$) &        &          &          &
        &          & 	     & \\ 
\noalign{\smallskip}
\hline
\noalign{\smallskip}
A     & 150 &  3000 & 6$\times 10^{16}$ & 1.0 & --- & ---  & ---               & --- & 0.9 & 0.8 & 0.6 & 43 & 21 & 21 \\
B     & 150 & 10000 & 3$\times 10^{17}$ & 1.0 & --- & ---  & ---               & --- & 0.7 & 1.0 & 0.9 & 35 & 13 &  9 \\
Total & 150 & 10000 & 1$\times 10^{16}$ & 0.7 & 150 & 1000 & 3$\times 10^{16}$ & 0.3 & 0.6 & 1.1 & 0.8 & 41 & 38 & 42 \\
Total & 150 & 10000 & 6$\times 10^{15}$ & 0.6 & 150 &  500 & 1$\times 10^{16}$ & 0.4 & 0.6 & 1.1 & 0.8 & 45 & 49 & 40 \\
\noalign{\smallskip}
\hline
\end{tabular}
\end{flushleft}
Notes: a. Line Ratios: $R_{i2} = \co(J$=$i$--$i$-1)/$\co(J$=2--1); 
$R_{j}$ = $\co$/$\13co$ ($J=j$--$j$-1); values of $R_{j}$ are calculated 
for a $\co/\13co$ abundance ratio of 60. Because of low optical 
depths, $R_{j}$ values scale linearly with assumed abundance ratio
\end{table*}

The characteristics of the Hubble~V environment (strong local radiation 
fields, low metallicity and relatively little radiation shielding) are
typically those of a strongly photon-dominated region (PDR). The data
collected in this paper provide good diagnostics for the PDR
properties of Hubble~V. The observed ratios log $I$([CII])/$I$(FIR) = 
-2.5 and log $I$(CO J=1--0)/$I$(FIR) = -6.8 can be used as analytical 
tools in the PDR models calculated by Kaufman et al. (1999 -- notably
their Figs.~17 and 1). These 
ratios apply to low-metallicity surroundings suffering moderate 
slab extinctions $A_{\rm V} \approx$ 3, radiation fields $G_{\rm o}$ = 
100 -- 300 and high densities $N_{\rm o} \approx 10^{4}$; 
$n_{\rm CII} \leq n_{\rm CO}$.  Kinetic temperatures should be of the 
order of $T_{\rm kin} \approx$ 150 K. 

Further details are provided by radiative transfer models (Jansen 1995; 
Jansen et al. 1994) which yield model CO line intensities as a function 
of gas kinetic temperature, $\h2$ density and CO column density per 
unit velocity. These model line intensities are coupled to actually 
observed values by a further parameter, the beam filling factor 
$f_{\rm CO}$. Thus, the four free parameters are essentially constrained 
by the four observed $\co$ transitions. Additional weak constraints are 
provided by the $\13co$ measurements. However, the line ratios are not 
uniqely diagnostic for the possible combinations of input parameters. 
For instance, non-identical but nevertheless very similar line ratios 
are produced by simultaneously reducing $\h2$ density and increasing 
kinetic temperature, with only little tweaking of CO column densities 
required. As the the actual measurements suffer from finite (and not 
very small) errors, they do not yield unique solutions.

It turns out that the line ratios of the positions A and B listed in 
Table 5 each yield two different solutions: a gas of relatively low 
temperature $T_{\rm kin}$ = 30 K but with high densities $n(\h2) = 
10^{4} - 10^{5}$, or a significantly hotter gas $T_{\rm kin}$ = 100 -- 
150 K of lower density $n(\h2) = 3\times 10^{3} - 10^{4}$. As shown at 
the beginning of this section, the latter solutions must be prefered
in view of the relative intensities of CO, [CII] and FIR emission. 
The line ratios for positions A and B could be fitted with a single gas
component, but this is not possible for the integrated emission of 
Hubble~V. If we assume that the ratios listed in Table 5 are actually
lower limits, the severity of this problem only increases.

Accordingly, we have modelled this emission with two separate components. 
Because the consequent doubling of free parameters brings their number to a 
total of eight, in excess of the number of independent measurements, this 
will always yield a set of non-unique solutions. As before, we reject
low temperature/high density solutions, thus reducing the range of possible 
solutions. We find the hottest and densest component to be tightly 
constrained by the observations, with $T_{\rm kin}$ = 150 K, 
$n(\h2) = 10^{4} \cc$ and $N({\rm CO})/$d$V$ = 0.6--1.0 $\times 10^{16} \cm2$. 
However, the more tenuous component is only weakly constrained: in 
principle, $\h2$ densities can be anywhere between 500 $\cc$ and 3000 $\cc$, 
kinetic temperatures are anywhere between 150 K (for the lower densities) 
and 20 K (for the higher densities). We have therefore further reduced 
the range of possible solutions by retaining only those cases where the 
more tenuous component is not cooler than the denser component, as we 
consider this to be physically more plausible than the reverse. We 
have listed the the resulting representative model solutions in Table 6. 
It is important to note that more accurate determination of $\co$ 
intensities would not greatly improve the situation as the various 
model-predicted line ratios are virtually identical. Better determination 
of $\13co$ intensities would help, but only if it were done very accurately 
as even here differences between the various solutions are not very great 
(see Table 6).

\subsection{Excited $\h2$ and other molecular species}

Emission by $\h2$ has been detected in its (1--0) S(1) transition with
a very low surface brightness of $4 \times 10^{-6}$ erg s$^{-1}
\cm2$ sr$^{-1}$ averaged over a relatively large aperture of 19.6$''$. 
However, the lack of a detection in the $\h2$ image at arcsec resolutions
(Sect. 2.3) above a level of $4 \times 10^{-5}$ erg s$^{-1} \cm2$ sr$^{-1}$ 
suggests that this $\h2$ emission is relatively widespread, and does not
contain high-contrast structure. This, in turn, suggests that the $\h2$ 
emission is dominated by UV rather than by shock excitation as is the case 
for HII regions in the Magellanic Clouds (Israel $\&$ Koornneef 1991). 
The values for Hubble~V are also quite reminiscent of those pertaining
to NGC~604, the much brighter first-ranked HII region in M~33. Towards
that complex, Israel et al. (1989) measured emission from radiatively
excited (fluorescent) $\h2$ with an almost identical mean surface brightness.

In Table 5 we have also compared the millimeter and submillimeter line 
ratios determined for Hubble~V with those found for sources in the 
Magellanic Clouds and the starburst galaxy He~2-10. Generally, Hubble~V 
appears to be hotter than most sources in the LMC and the SMC, and to 
suffer more from CO photo-dissociation. The line ratios for the fairly 
extreme 30 Doradus PDR and the starburst core of He~2-10 come closest to 
those of Hubble~V. It is of particular interest to note that in Hubble~V, 
the $J$=1--0 HCO$^{+}$ line is much stronger than the $J$=1--0 $\13co$ 
line (see Fig.~\ref{molprofiles}) as usually the reverse is the case. If 
we assume HCO$^{+}$ to be at the same high temperature $T_{\rm kin}$ = 
150 K as the CO, we obtain an HCO$^{+}$/CO abundance of 10$^{-3}$, which 
is much higher than found for the Magellanic regions (Chin et al. 1997, 
1998; Heikkil\"a et al. 1999). Because CO suffers from relatively intense 
photo-destruction, the HCO$^{+}$ overabundance is less extreme when 
related to $\h2$. In any case, the relatively high intensity of 
HCO$^{+}$ emission is in line with conclusions by Johansson et al. 
(1994) and Heikkil\"a et al. (1999) that this line is enhanced in 
active star formation regions. The possible detections of the CS 
$J$=3--2 and H$_{2}$CO 2$_{1,1}$--1$_{1,0}$ transitions are so 
marginal that no useful conclusions can be based on them.

\subsection{Mass of the Hubble~V complex}

The amount of carbon monoxide associated with Hubble~V is not very
large. First, the emission has a low optical depth. In fact, the high-density
component in the last two rows of Table 6 is optically thin in all four
transitions observed. Second, its surface filling factor is low. 
For the total source, surface filling factors are of order 0.3 and 0.17 for
the dense and tenuous components respectively. Third, the extent of CO
emission is small, so that e.g. the $J$=1--0 $\co$ measurements suffer 
also from beam dilution as the 43$''$ beam is larger than the CO
source. 

Although the peak of the [CII] emission coincides, within the limits 
imposed by a limited resolution, with the HII/CO complex, this emission 
extends well beyond the boundaries of the complex and follows the HI 
distribution (Figs.~\ref{ciimap}, ~\ref{himaps}). Because the HI column 
density towards Hubble~V and its surroundings is of order $N(HI) \approx 
1-2 \times 10^{21} \cm2$, it is likely that the extent of molecular
hydrogen also significantly exceeds that of CO. In particular at the 
high radiation field intensities characterizing Hubble~V, the expected 
(Kaufman et al. 1999) and observed intensities imply an overall dilution
of the [CII] emission due to incomplete surface and beam filling by
a factor of 5 to 10. From the radiative transfer models, we find that in
Hubble~V, C$^{+}$ column densities on average exceed those of CO by a 
large factor of 20$\pm$10. Indeed, the ratio $F_{\rm CII}/F_{\rm CO} = 
3\pm 1 \times 10^{4}$ is identical to the high ratio that Israel et al. 
(1996b) found in the LMC complex N~160 which they considered to be in an 
advanced stage of CO destruction. 

If we neglect possible contributions by C$^{\rm o}$, the models provide 
total carbon column densities which can be converted to total hydrogen 
column densities, provided the [C]/[H] abundance is known. This has not 
yet been measured for Hubble~V, but we may use the data collected by 
Garnett et al. (1999, notably their Fig. 4) to estimate [C]/[H] from 
the known oxygen abundance [O]/[H]. If we furthermore estimate that in 
relatively cool molecular environments about 70$\%$ of all carbon will 
be locked up in dust grains (i.e. the depletion factor $\delta_{\rm C}$ 
= 0.3), we find a gas-phase abundance ratio $\delta_{\rm C} \times$
[C]/[H] $\approx$ 1.8 $\times 10^{-5}$. From this, we find for the 
Hubble~V complex a molecular mass $M(\h2) = (6\pm 3) \times 10^{5}$ 
M$_{\odot}$ and a total gas mass, including HI and He, of $M_{gas} = 
(10\pm 5) \times 10^{5}$ M$_{\odot}$. 

These masses are an order of magnitude greater than the {\it upper limit} 
to the ionized hydrogen mass $M(HII) \leq 7 \times 10^{4}$ M$_{\odot}$ 
derived from the radio flux-densities in Sect.~2.1 under the assumption 
of a homogeneous gas distribution (cf. Mezger $\&$ Henderson 1967) 
and also much larger than the total mass $M(O) \approx 2 \times 10^{3}$ 
M$_{\odot}$ estimated by Wilson (1992a) contained in stars with mass 
greater than 15 M$_{\odot}$ (i.e. earlier than spectral type B0). In Sect. 
3.2 we surmised that a similar number of stars may have escaped attention 
by having high extinctions. Moreover, if the stellar ensemble associated 
with HUubble~V is associared with a Miller-Scalo (1979) initial mass 
function, nonionizing stars may increase the total stellar mass by a 
factor of about 20. Thus, the total mass of all embedded stars will 
be of the order of $10^{5}$ M$_{\odot}$, i.e. an order of magnitude 
less than the total neutral gas mass.

The molecular results imply a CO-to-$\h2$ conversion factor of $X$ = 
54$\pm$27 $\times 10^{20} \cm2 (\kkms)^{-1}$. The fact that this result is
{\it identical} to that obtained in Paper II is no doubt fortuitous. This
independently derived result does confirm, however, the high value of
the $X$-factor calculated there.

\section{Conclusions}

\begin{enumerate}
\item We have presented maps of the $\co$ emission in four transitions
associated with the extragalactic HII region complex Hubble~V. 
The extent of the CO emission is rather limited, and
comparable to the extent of the ionized gas forming the HII region.
\item We have also measured the integrated emission from the complex
in transitions of HCO$^{+}$ and marginally detected $J$=1--0 $\13co$,
$J$=3--2 CS and H$_{2}$CO. The complex and its surroundings were mapped
in C$^{+}$. The resulting various line ratios show that most CO in Hubble~V 
is optically thin in at least the lower observed transitions.
\item The Hubble~V complex has all the characteristics of a relatively
extreme photon-dominated region (PDR), in which a severely eroded CO core 
exists in a significantly larger cloud of $\h2$ and HI. In such an extreme 
PDR, cloud parameters appear better represented by the properties
of the dissociation product carbon, traced by [CII] emission, than by 
the remnant CO. 
\item The average reddening of the HII region and the visible OB
associations is $E(B-V)$ = 0.50-0.65 mag. However, the molecular cloud 
obscures most of the stars responsible for the excitation of Hubble~V 
at optical wavelengths although they can be seen in the near-IR $K$-band.
\item Comparison of the observations to chemical and radiative
transfer models indicates that the CO occurs in high-density (typically
$n(\h2) = 10^{4} \cc$) parts of the molecular cloud complex at high
temperatures ($T_{\rm kin} \approx$150 K). Although the space density of
molecular gas containing C$^{+}$ may be lower than that traced by CO
emission, the former has much higher column densities.
\item The mass of the whole complex is of the order of 10$^{6}$ M$_{\odot}$.
Slightly less than two thirds of this mass is in the form of molecular gas,
and the remainder is mostly neutral atomic gas. The ionized gas and the 
embedded stars account for typically $15\%$ of the total mass. The CO
molecule is rather efficiently destroyed in Hubble~V, even though the 
efficiency of star formation is not particularly high.
\item The weakness of CO emission, and its small spatial extent do not 
imply a similar paucity of $\h2$ gas. The relatively low metallicity of
Hubble~V provides insufficient shielding for CO in most of the
complex; the strongly selfshielding $\h2$ is only little affected
by strong but not extremely intense radiation fields.
As a consequence, Hubble~V is characterized by a rather high CO-to$\h2$
conversion factor $X \approx 5 \times 10^{21} \cm2 \kms^{-1}$, as
indeed surmised earlier.
\end{enumerate}

\acknowledgements

It is a pleasure to thank the operating personnel of the SEST and the JCMT
for their support, and Ute Lisenfeld for conducting the IRAM 30 m service
observations. The [CII] measurement was obtained in a program of
measuring various Magellanic Cloud HII regions, involving also Phil
Maloney, Gordon Stacey, Sue Madden and Albrecht Poglitsch. Some of the
near-infrared observations of $\h2$ and Br$\gamma$ were likewise obtained 
within the framework of a different observing program with Paul van der Werf.
Bob O'Dell and Luciana Bianchi kindly permitted use of the NASA/ESA/STScI
HST image of Hubble~V based on their data. RJR was supported by the 
Independent Research and Development program at The Aerospace Corporation.

\end{document}